\documentclass[aps,pre,showpacs,showkeys,amsmath,amsfonts,amssymb,preprint]{revtex4}

\newcommand{\beq}{\begin{equation}}
\newcommand{\eeq}{\end{equation}}
\newcommand{\bea}{\begin{eqnarray}}
\newcommand{\eea}{\end{eqnarray}}
\newtheorem{theorem}{Theorem}[section]

\input epsf

\begin{document}

\title{Phase transitions and topology changes in configuration space}

\author{Lapo Casetti}
\email{casetti@fi.infn.it}
\affiliation{Istituto Nazionale per la Fisica della Materia (INFM), Unit\`a di
Ricerca di Firenze, via G.~Sansone 1, I-50019 Sesto Fiorentino (FI), Italy}

\author{Marco Pettini}
\email{pettini@arcetri.astro.it}
\altaffiliation[Also at: ]{INFM, UdR Firenze, and INFN, Sezione di Firenze, 
via G.~Sansone 1, I-50019 Sesto Fiorentino (FI), Italy.}
\affiliation{Istituto Nazionale di Astrofisica,
Osservatorio Astrofisico di Arcetri, Largo E.~Fermi 5,
I-50125 Firenze, Italy} 

\author{E.~G.~D.~Cohen} 
\email{egdc@rockefeller.edu} 
\affiliation{The Rockefeller University, Box 28, 1230 York Avenue, New York, 
New York  10021 - 6399} 

\date{\today}

\begin{abstract}
The relation between thermodynamic phase transitions in classical systems
and topological changes in their configuration space
is discussed for two physical models and contains the first 
{\it exact analytic} computation of 
a topologic invariant (the Euler characteristic) of certain submanifolds
in the configuration space of two physical models.
The models are the mean-field $XY$ model and the one-dimensional $XY$
model with nearest-neighbor interactions. The former model undergoes a
second-order phase transition at a finite critical temperature while the 
latter has no phase transitions.
The computation of this topologic invariant is performed within the framework
of Morse theory.
In both
models topology changes in configuration space are present as the potential
energy is varied; however, in the
mean-field model there is a particularly ``strong'' topology change,  
corresponding to a big jump in the Euler characteristic, connected with 
the phase transition, which is absent in the one-dimensional model with
no phase transition. The comparison between the two models has two major
consequences: {\it i)} it lends new and strong support to a recently proposed
topological approach to the study  of phase transitions; {\it ii)} it allows 
us to conjecture which particular topology changes could entail a phase 
transition in general. 
We also discuss a simplified illustrative model of the topology changes 
connected to phase transitions using of two-dimensional
surfaces, and a possible direct connection between topological invariants and 
thermodynamic quantities.

\end{abstract}

\pacs{75.10.Hk; 02.40.-k; 05.70.Fh; 64.60.Cn}

\keywords{Phase transitions; topology; configuration space; mean-field models}

\maketitle

\section{Introduction}

One can wonder whether the current mathematical description of thermodynamic 
phase transitions (based on the loss of analyticity of thermodynamic 
observables \cite{LeeYang,Ruelle,Sinai}) is the ultimate possible one, 
or whether a reduction to a deeper mathematical level is possible. 

Besides a purely theoretical motivation, there are other
reasons for thinking of such a possibility. Among the others, we mention the
growing experimental evidence that phase transitions occur in very small $N$
systems, like nuclear clusters as well as atomic and molecular clusters, 
in nano and mesoscopic systems, in polymers and proteins, in very
small drops of quantum fluids (BEC, superfluids and superconductors).

Moreover, new mathematical characterizations of thermodynamic phase 
transitions could well be of interest for the treatment of other 
important topics in statistical physics, as is the case of amorphous 
and disordered systems (like glasses and spin-glasses), or to incorporate 
also first-order phase transitions. A different attempt then is discussed
here has been made in macroscopic parameter space instead of in microscopic
phase space in Ref.\ \cite{gross}.

In a number of recent papers \cite{cccp,franzosi,xymf,phi4,physrep,CSCP} 
a proposal has been put forward for a new mathematical approach to the
study of phase transitions. This applies to physical systems described by 
continuous variables -- $q_i$ and $p_i$, $i = 1,\ldots,N$ -- entering a 
standard Hamiltonian
\begin{equation}
{\cal H} = \frac{1}{2}\sum_{i=1}^N p_i^2 + V(q_1,\ldots,q_N)~,
\label{ham_standard}
\end{equation}
where $V(q_1,\ldots,q_N)$ is the potential
energy. 
The main issue of this new 
approach is a {\em Topological Hypothesis} (TH). The content of the TH is 
that at their deepest level phase
transitions of a system are due to a change of the topology of suitable 
submanifolds in its configuration space.
More precisely, once the microscopic interaction potential 
$V(q_1,\ldots,q_N)$ is given, the
configuration space of the system is automatically foliated into the family
$\{ \Sigma_v\}_{v\in{\Bbb R}}$ of equipotential hypersurfaces independently of
any statistical measure we may wish to use.
Now, from standard statistical mechanical arguments we know that the larger 
the number $N$ of particles, the closer to some $\Sigma_v$ are the
microstates which significantly contribute to the statistical averages 
of thermodynamic observables. At large $N$, and at any given value of the
inverse temperature $\beta$, the effective support of the canonical measure
is narrowed very closely to a single hypersurface
$\Sigma_v \equiv \{ q \in {\Bbb R}^N \vert
 V(q) = v \} \subset {\Bbb R}^N$, with $v$ a suitable function of $\beta$.

Now, the TH consists in assuming that some suitable
change of the topology  of the $\{ \Sigma_v\}$, or equivalently, at large $N$,
of the 
submanifolds $M_v=\{ q\in{\Bbb R}^N\vert V(q)\leq v\}$ (the manifolds $\Sigma_v$
are the boundaries of the $M_v$, i.e., $\Sigma_v =
\partial M_v$),  
occurring at some $v_c=v_c(\beta_c)$ (or $v_c=v_c(E_c)$),  is the deep
origin of the singular behavior of thermodynamic observables  at a
phase transition; (by change of topology we mean that  $\{ \Sigma_v\}_{v<v_c}$
are {\it not diffeomorphic}\footnote{A diffeomorphism is a smooth map between
two differentiable manifolds which is invertible together with its derivatives.} 
to the $\{ \Sigma_v\}_{v>v_c}$, or equivalently that $\{ M_v\}_{v<v_c}$
are {\it not diffeomorphic} 
to the $\{ M_v\}_{v>v_c}$). In the following of the paper we shall consider the
topology changes of the $M_v$, because these are naturally investigated using
Morse theory.

The present paper contributes to this new approach to phase 
transitions with a crucial step forward. This consists of   
an {\it exact analytic} treatment of
the topology changes in the configuration space of the mean-field $XY$ model
and of their relation to the thermodynamic phase transition, reported
in Sec.~\ref{sec_MF}. In Sec.~\ref{sec_1d} we present
an analogous treatment of a closely
related model, the one-dimensional $XY$ model with nearest-neighbors
interactions, which does not have any thermodynamic phase transition, and we
compare the results with those obtained for the mean-field case. In
Sec.~\ref{sec_model} we present and discuss a geometric model built in terms of
two-dimensional surfaces which should help the intuition in understanding some
of the aspects of the much more complex topology changes in $N$-dimensional
configuration spaces. Sec.~\ref{sec_comments} is devoted to some final remarks
and to the discussion of possible future developments. The paper is completed
by Appendix \ref{app_proof} which 
is devoted to the proof of some crucial estimates used in the main text.

\section{The mean-field $XY$ model}
\label{sec_MF}

In this section we give a complete analytical characterization of the
topological changes in the configuration space of a model with long-range
interactions, the mean-field $XY$ model, including the one related to the phase
transition. This is, to our knowledge, the first complete analytical
characterization of a topological change in configuration space
 related to a thermodynamic phase 
transition: a short account of this result has already been given in
\cite{euler}. For a family of submanifolds of
the configuration space defined by the potential energy, 
we are able -- for the first time for a model of physical relevance -- 
to determine completely, in a constructive way, the
topology and to 
give analytical estimates of the Betti numbers, i.e.,
of fundamental topological invariants\footnote{The Betti numbers $b_k$ of
a differentiable manifold $M$ are the dimensions of the de Rham's cohomology 
vector spaces ${\Bbb H}^k(M;{\Bbb R})$, that is the vector 
spaces of closed differential $k$-forms modulo the exact forms of the same 
order. These are diffeomorphism invariants.}. This allows us 
to compute exactly a topological invariant: the Euler characteristic, 
which can be defined as a combination of the Betti numbers \cite{nakahara}. 

The mean-field $XY$ model 
\cite{AntoniRuffo} is defined by a Hamiltonian of the class
(\ref{ham_standard}) where the potential energy function is
\begin{equation}
V(\varphi) = \frac{J}{2N}\sum_{i,j=1}^N 
\left[ 1 - \cos(\varphi_i - \varphi_j)\right] -h\sum_{i=1}^N \cos\varphi_i ~.
\label{V}
\end{equation}
Here $\varphi_i \in [0,2\pi]$ is the rotation angle of the $i$-th rotator 
and $h$ is an external field. Defining at each site $i$ a 
classical spin vector ${\bf s}_i = (\cos\varphi_i,\sin\varphi_i)$ the model 
describes a planar (XY) Heisenberg system with interactions of equal strength 
among all the spins. We consider only the ferromagnetic case $J >0$; 
for the sake of simplicity, we set $J=1$. The equilibrium statistical mechanics 
of this system is exactly described, in the thermodynamic limit, by mean-field 
theory. In the limit $h \to 0$, the system has a continuous 
phase transition, with classical critical exponents, at $T_c = 1/2$, or 
$\varepsilon_c = 3/4$, 
where $\varepsilon = E/N$ is the energy per particle \cite{AntoniRuffo}. 

Defining the magnetization vector ${\bf m}$ as 
${\bf m} = \left(m_x,m_y\right)~$, where
\begin{eqnarray}
m_x & = & \frac{1}{N}\sum_{i=1}^N \cos\varphi_i~, \\
m_y & = & \frac{1}{N}\sum_{i=1}^N \sin\varphi_i~,
\end{eqnarray}
the potential energy $V$ can be written as a function of $\bf m$ as follows:
\begin{equation}
V(\varphi) = V(m_x,m_y) = \frac{N}{2} (1 - m_x^2 - m_y^2) 
- hN\, m_x~.
\label{V(m)}
\end{equation}
The range of values of the potential energy per particle, 
${\cal V} = {V}/{N}~$, is then 
\begin{equation}
-h \leq {\cal V} \leq \frac{1}{2} + \frac{h^2}{2}~.
\end{equation}

The configuration space $M$ of the model is an $N$-dimensional torus, being
parametrized by $N$ angles. We want to investigate the topology of the 
following family of submanifolds of $M$,
\begin{equation}
M_v = {\cal V}^{-1} (-\infty,v] = 
\{ \varphi \in M : {\cal V}(\varphi) \leq v\}~,
\label{submanifolds}
\end{equation}
i.e., each $M_v$ is the set $\{\varphi_i\}_{i=1}^N$ such that the potential 
energy per particle does not exceed a given value $v$: this is
the same as the $M_v = M_{E-K}$ defined above 
($v$ has been rescaled by $\frac{1}{N}$ because we choose ${\cal V} = V/N$ as a
Morse function
in order to make the comparison of systems with different $N$ easier).  
As $v$ is increased from $-\infty$ to $+\infty$, this
family covers successively the whole manifold $M$ ($M_v \equiv \emptyset$ 
when $v< -h$) .

According to Morse theory \cite{milnor}, 
topology changes of $M_v$ can occur only in
correspondence with critical points of $\cal V$, i.e., those points where the
differential of $\cal V$ vanishes. This immediately implies that
no topology changes can occur when $v > 1/2 + h^2/2$, i.e., all the $M_v$'s 
with $v > 1/2 + h^2/2$ 
must be diffeomorphic to the whole $M$, that is, they must be $N$-tori.
Moreover, if $\cal V$ is a Morse function (i.e., it has only non-degenerate
critical points) then topology changes of $M_v$ are actually in one-to-one
correspondence with critical points of $\cal V$, and they can be characterized
completely. At any critical level of $\cal V$
the
topology of $M_v$ changes in a way 
completely determined by the local properties of the Morse function: 
a $k$-handle $H^{(k)}$ is attached\footnote{A $k$-handle $H^{(k)}$ in $n$ 
dimensions 
$(0\leq k\leq n)$ is the product of two disks, one $k$-dimensional, $D^k$,
and one $(n-k)$-dimensional, $D^{(n-k)}$: $~ H^{(k)}=D^k\times D^{(n-k)}$.}
 \cite{milnor},  where $k$ is
the {\em index} of the critical point, i.e., the number of negative eigenvalues
of the Hessian matrix of $\cal V$ at this point. 
Notice that if there are $m>1$ critical
points on the same critical level, with indices $k_1,\ldots,k_m$, then the
topology change is made by attaching $m$ disjoint handles
$H^{(k_1)},\ldots,H^{(k_m)}$.
This way, by increasing $v$, the topology of the 
full configuration space $M$ can be 
constructed sequentially from the $M_v$.
Knowing the index of all the critical points below a given level $v$, we can 
obtain {\it exactly} the Euler characteristic of the manifolds $M_v$, defined 
by
\begin{equation}
\chi (M_v) = \sum_{k = 0}^N (-1)^k \mu_k(M_v)~,
\label{chi_morse}
\end{equation}
where the {\em Morse number} $\mu_k$ is the number of critical points of $\cal V$
which have index $k$  \cite{milnor}. 
The Euler characteristic $\chi$ is a
{\em topological invariant} (i.e it is not affected by a diffeomorphic
deformation of $M_v$): any change in $\chi(M_v)$ implies a
topology change in the $M_v$.
It will turn out that as long as $h > 0$, 
$\cal V$ is indeed a Morse function at least
in the interval $-h \leq v < 1/2 + h^2/2$, while the maximum value $v = 1/2 +
h^2/2$ may be pathological in that it may correspond to a 
critical level with degenerate critical points. However, as we shall show in the
following, we shall be able to extend our analysis also to this last critical
level.

Thus, in order to detect and characterize topological 
changes in $M_v$ we have to find the 
critical points and the critical 
values of ${\cal V}$, which means solving the equations
\begin{equation}
\frac{\partial {\cal V}(\varphi)}{\partial \varphi_i} = 0~, 
\qquad i = 1,\ldots,N~,
\label{crit_eqs}
\end{equation}
and to compute the indices of {\em all} the critical points of $\cal V$,
i.e., the number of negative eigenvalues of its Hessian
\begin{equation}
H_{ij} = \frac{\partial^2 {\cal V}}{\partial \varphi_i \partial \varphi_j}  
\qquad i,j = 1,\ldots, N\,.
\label{hess}
\end{equation} 

Taking advantage of Eq.\ (\ref{V(m)}), we can rewrite the equations
(\ref{crit_eqs}) as
\begin{equation}
(m_x + h) \sin\varphi_i - m_y \cos\varphi_i = 0 ~ , 
\qquad i = 1,\ldots,N~.
\label{crit_eqs_i}
\end{equation}
As long as $(m_x + h)$ and $m_y$ are not simultaneously zero 
(the violation of this
condition is possible only on the level $v = 1/2 + h^2/2$), the solutions of
Eqs.\ (\ref{crit_eqs_i}) are all configurations in which the angles are
either $0$ or $\pi$. In particular, the configuration
\begin{equation}
\varphi_i = 0 \qquad \forall\, i
\end{equation}
is the absolute minimum of $\cal V$, while all the other configurations
correspond to a value of $v$ which depends only on the number of angles which
are equal to $\pi$. If we denote with $n_\pi$ this number, we have
that the $N$ critical values are:
\begin{equation}
v(n_\pi) = \frac{1}{2}\left[1 - \frac{1}{N^2}\left(N - 2n_\pi\right)^2\right] 
- \frac{h}{N}\left(N - 2n_\pi\right)\,. 
\label{v(npi)}
\end{equation}
Inverting this relation yields $n_\pi$ as a function of the level value $v$:
\begin{equation}
n_\pi(v) = {\mbox{\rm int}}\, \left[\frac{1 + h}{2} N \pm 
\frac{N}{2} \sqrt{h^2 - 2 \left(v - \frac{1}{2} \right)} \right]\,,
\label{n(v)}
\end{equation}
where ${\mbox{\rm int}}\,[a]$ stands for the integer part of $a$.
We can also compute the number $C(n_\pi)$ of critical points having a 
given $n_\pi$, which is the number of distinct binary strings of length $N$
having $n_\pi$ occurrences of one of the symbols, which is given by
the binomial coefficient
\begin{equation}
C(n_\pi) = {N \choose n_\pi} = 
\frac{N!}{n_\pi ! \left(N - n_\pi \right)!} ~.
\label{n_cp}
\end{equation}

We have thus shown that as $v$ changes from its minimum $-h$ 
(corresponding to $n_\pi = 0$) to  
$\frac{1}{2}$ (corresponding to $n_\pi = \frac{N}{2}$)
the manifolds $M_v$ undergo a sequence of topology
changes at the $N$ critical values $v(n_\pi)$ given by Eq.~(\ref{v(npi)}). 
We expect that there is another topology change located
at the last (maximum) critical value,
\begin{equation}  
v_c = \frac{1}{2} +
\frac{h^2}{2}~.
\label{v_c}
\end{equation}
However, the above argument does not prove this, 
since the critical points of $\cal V$ corresponding to this critical 
level may be degenerate. This, because the solutions of the two equations in 
$N$ variables
$m_x = m_y = 0$ need not to be isolated,  
so that then on this level, $\cal V$ would not be a proper Morse
function. Then a critical value $v_c$ is still a necessary condition for the
existence of a topology change, but it is no longer sufficient \cite{Palais}.  
However, as already argued in Refs.~\cite{xymf,physrep}, 
it is just this topology change occurring at $v_c$ given in
Eq.~(\ref{v_c}), which is 
related to the thermodynamic phase transition of the mean-field XY model. 
For, the temperature $T$, the 
energy per particle $\varepsilon$ and the average potential energy per 
particle $u = \langle {\cal V} \rangle$ obey, 
in the thermodynamic limit, the equation 
\begin{equation}
2\varepsilon = T + 2u(T)~;
\end{equation}
substituting in this equation
the values of the critical energy per particle  and of the critical
temperature we get  
\begin{equation}
u_c = u(T_c) = 1/2~.
\end{equation} 
as $h \to 0$, $v_c \to \frac{1}{2}$, 
so that $v_c = u_c~$.
Thus a topology change in the family of manifolds $M_v$ 
occurring at this $v_c$, 
where $v_c$ is {\em independent} of $N$, is connected with the
phase transition in the limit $N\to\infty$, and $h \to 0$,
when indeed thermodynamic phase transitions are usually defined, at least in the
canonical ensemble. 

We still have to prove that
a topology change at $v_c$ actually exists. 
To carry out this proof, we will use Morse
theory to characterize completely all the other topology changes, occurring at
$v < v_c$: this, together with the knowledge that at $v > v_c$ the manifold
$M_v$ must be an $N$-torus, will allow us not only to prove that a topology
change at $v_c$  must actually occur, but also to understand in which way it
is different from the other topology changes, i.e., 
those occurring at $0 \leq v < v_c$. 
Morse theory allows a complete
characterization of the topology changes occurring in the $M_v$'s if the {\em
indices} of the critical points of $\cal V$ are known. In order to determine 
the indices of the critical points (that is the number of negative eigenvalues
of the Hessian of  $\cal V$ at the critical point) we proceed as follows.
Since the diagonal elements of the Hessian are
\begin{equation}
H_{ii} = d_i = \frac{1}{N}\left[(m_x + h) \cos\varphi_i + m_y
\sin\varphi_i\right] - \frac{1}{N^2}~,
\end{equation}
and the off-diagonal elements are
\begin{equation}
H_{ij} = - \frac{1}{N^2}\left(\sin\varphi_i\sin\varphi_j +
\cos\varphi_i\cos\varphi_j \right)\,,
\label{nondiag}
\end{equation}
one can write the Hessian as the sum of a diagonal matrix $D$ whose
nonzero elements are
\begin{equation}
\delta_i = \frac{1}{N}\left[(m_x + h) \cos\varphi_i + m_y
\sin\varphi_i\right] ~, \qquad i = 1,\ldots,N~,
\label{deltas}
\end{equation}
and of a matrix $B$ whose elements are just the $H_{ij}$ given in
Eq.~(\ref{nondiag}), also for $i=j$ (the diagonal elements being $-1/N^2$).
Since the ratio between the elements of $B$ and those of $D$ 
is ${\cal O} (1/N)$,
one would expect at first sight that only the diagonal elements survive when
$N \gg 1$, so that the Hessian approaches a diagonal matrix equal to $D$. 
However, this is not, in principle, necessarily true: one cannot immediately 
say that at large $N$ the eigenvalues of the Hessian are the $\delta$'s 
given in
Eq.~(\ref{deltas}) plus a correction vanishing as $N \to \infty$, because the
number of elements of $B$ is $N^2$, so that the contribution of the matrix $B$
to the eigenvalues of the Hessian does {\em not}, in general, vanish at large
$N$.  That nevertheless the argument for this crucial point is correct
in this special case is shown in Appendix \ref{app_proof}, and is
due to the particular structure of the
matrix $B$. The latter is of rank one, and 
one can prove then that at the critical points of $\cal V$ 
the number of negative
eigenvalues of $H$ equals the number of negative 
diagonal elements $\delta$ $\pm 1$, 
so that as $N \gg 1$ we can conveniently approximate the index of the
critical points with the number of negative $\delta$'s at $x$,
\begin{equation}
{\rm index}~(x) \simeq \#(\delta_i < 0)~.
\label{estimate}
\end{equation}

At a given critical point, 
with given $n_\pi$,  where the $x$-component of the magnetization vector is 
\begin{equation}
m_x = 1 - \frac{2n_\pi}{N}~,
\end{equation} 
so that $m_x >  0$ (resp.~$<0$) if $n_\pi \leq \frac{N}{2}$ (resp.~$ >
\frac{N}{2}$), 
the eigenvalues of $D$ are 
\begin{subequations}
\begin{eqnarray}
\delta_i & = & m_x + h \qquad \qquad i  =  1,\ldots,N - n_\pi\,; \\
\delta_i & = & -(m_x + h) \qquad ~ i  =  N - n_\pi + 1,\ldots,N\,.
\end{eqnarray}
\end{subequations}
Then, if the external field $h$ is sufficiently small, 
\begin{eqnarray}
\left(m_x + h
\right) & > & 0 \qquad  {\rm if}~ n_\pi \leq \frac{N}{2}~, \nonumber \\ 
\left(m_x + h
\right) & < & 0 \qquad  {\rm if}~ n_\pi > \frac{N}{2}~,
\end{eqnarray}
so that, denoting with ${\rm index}(n_\pi)$ the index of a critical point with
$n_\pi$ angles equal to $\pi$, we can write
\begin{subequations}
\begin{eqnarray}
{\rm index}(n_\pi) & = & n_\pi \qquad \qquad  {\rm if}~ n_\pi \leq
\hbox{$\frac{N}{2}$}~, \\
{\rm index}(n_\pi) & = & N - n_\pi \qquad  {\rm if}~ n_\pi > 
\hbox{$\frac{N}{2}$}~.
\end{eqnarray}
\label{index(n)}
\end{subequations}
From these equations combined with Eq.\ (\ref{n_cp}) one can obtain for the 
Morse numbers $\mu_k$, i.e., for the number of critical points of index $k$, as
a function of the level $v$, as long as $-h \leq v < 1/2 + h^2/2$ (i.e.,
excluding the limiting value $v = 1/2 + h^2/2$) the following expression:
\begin{equation}
\mu_k (v)  =  
{N \choose k} 
\left[1 - \Theta(k - n^{(-)}(v)) + \Theta(N - k - n^{(+)}(v)) \right] 
\, , 
\label{mu_k}
\end{equation}
where $\Theta(x)$ is the Heaviside theta function 
and $n^{(\pm)}(v)$ are the limits of the allowed $n_\pi$'s for a given
value of $v$, i.e., from Eq.\ (\ref{n(v)}),
\begin{equation}
n^{(\pm)}(v)  =  \frac{N}{2}\left[ 1 + h \pm 
\sqrt{h^2 - 2 \left(v - \frac{1}{2} \right)} \right]\,.
\end{equation}
We note that $0 \leq n_\pi^{(-)} \leq \frac{N}{2}$ and 
$ \frac{N}{2} + 1 \leq n_\pi^{(+)} \leq N$, so that Eq.\ (\ref{mu_k}) implies
\begin{equation}
\mu_k (v) = 0 \qquad \forall\,k > \frac{N}{2}~,
\end{equation}
i.e., no critical points with index larger than $N/2$ exist as long as 
$v < v_c = 1/2 + h^2/2$. 
This is the crucial observation to prove that a topology
change {\em must} occur at $v_c$. For, as the Betti numbers of a 
manifold are positive (or zero) 
numbers, using the Morse inequalities, which state that the Morse numbers 
are upper bounds of the
Betti numbers \cite{Palais}, i.e. 
\begin{equation}
b_k \leq \mu_k ~~{\rm for~~~}  k = 0,\ldots,N,
\label{morse_ineq}
\end{equation}
we can immediately conclude that as long as $v < v_c = 1/2 + h^2/2$
\begin{equation}
b_k (M_v) = 0 \qquad \forall\,k > \frac{N}{2}~.
\end{equation}
Thus, as $\frac{1}{2} \leq v < \frac{1}{2} + \frac{h^2}{2}$ 
the manifold is only ``half'' an $N$-torus, and
since we know that for $v > \frac{1}{2} + \frac{h^2}{2}$, $M_v$ {\em is} a
(full) $N$-torus, whose Betti numbers are 
\begin{equation}
b_k (\mathbb{T}^N) = {N \choose k} \qquad k = 0,1,\ldots,N\,,
\end{equation}
we conclude that at 
$v = v_c = \frac{1}{2} + \frac{h^2}{2}$ a topology change {\em must} occur, 
which involves the attaching of
$N\choose k$ different $k$-handles for each $k$ ranging from $\frac{N}{2} + 1$ 
to $N$ i.e. a change of ${\cal O}(N)$ of the number of Betti numbers. 

Let us remark that such a topology change not only exists: it is 
surely a ``big'' topology change, for all of a sudden, ``half'' an $N$-torus
becomes a full $N$-torus, via
the attaching for each different $k$ (ranging from $N/2 + 1$ to $N$)  of
$N\choose k$ $k$-handles. More precisely, a number of Betti numbers which is
${\cal O}(N)$ changes, and changes by amounts which are of the same
order as 
their maximum possible values. On the other hand, all the other topology 
changes
correspond to the attaching of handles of the same type (index), in fact, 
as long as $v < v_c$,
each critical level contains only critical points of the same index, and the
index grows with $v$, i.e., if $x_c$ and $x'_c$ are critical points and ${\cal
V}(x'_c) > {\cal V}(x_c)$, then $\mathrm{index}(x'_c) > \mathrm{index}(x_c)$.
The potential energy per degree of freedom $\cal V$ 
is a \emph{regular} Morse function (or a 
Morse-Smale function \cite{Palais}) as long as $v < v_c$, but this is no 
longer true as $v\geq v_c$; actually, as we have already observed, $\cal V$ 
could even be no longer a Morse function at all, 
because the level $v_c$ might contain degenerate
critical points. Nevertheless, as we have shown, this does not prevent
us from giving a complete analysis of the topology of the $M_v$'s for {\em
all} the values of $v$, since we can exploit our explicit knowledge of
the topology of the $M_v$'s for any $v > v_c$.

\begin{figure}
\epsfysize= 14 truecm 
\begin{center}
\mbox{\epsfbox{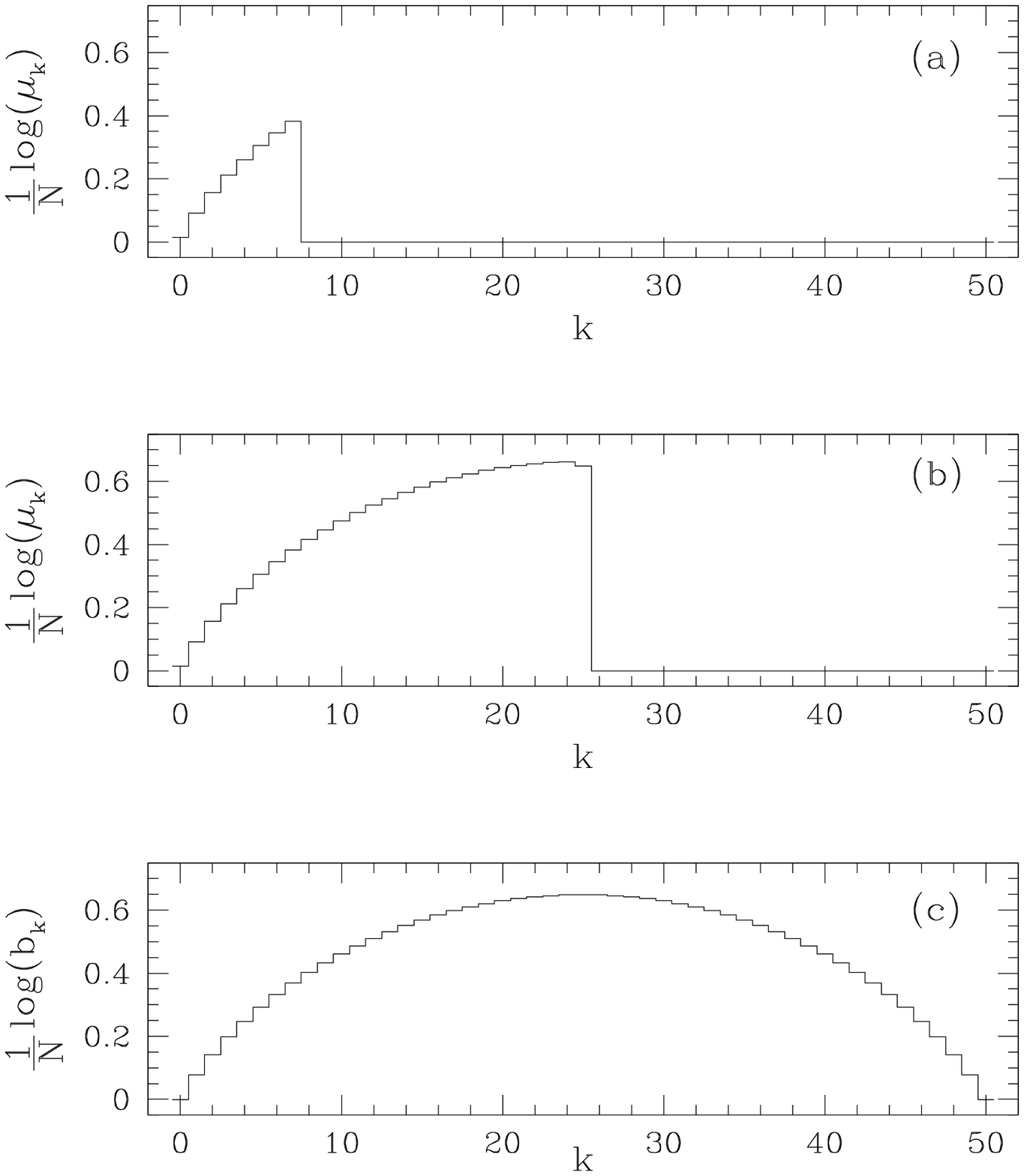}} 
\end{center}
\caption{Mean-field $XY$ model. 
$(a)$ Histogram of $\log(\mu_k(M_v))/N$ as a function of $k$ for
$v=1/4$; $(b)$ Histogram of $\log(\mu_k(M_v))/N$ as a function of $k$ for
$v=1/2$. In both cases $N =$
50 and $h = 0.01$.
$(c)$ For comparison, histogram of $\log(b_k(\protect\mathbb{T}^N))/N$ 
as a function of $k$ for
an $N$-torus $\protect\mathbb{T}^N$, with $N=50$, which is the lower bound of
$\log(\mu_k(M_v))/N$ for any $v \geq v_c$.}
\label{fig_morse_num}
\end{figure}

To illustrate what we have described so far, 
the Morse numbers $\mu_k$ are shown in Fig.~\ref{fig_morse_num} as a function 
of $k$ for two values of $v$, $v = \frac{1}{4}$, i.e., an intermediate value
between the minimum and the maximum of $\cal V$ -- shown in
Fig.~\ref{fig_morse_num} $(a)$ -- and $v = \frac{1}{2}$ -- shown in
Fig.~\ref{fig_morse_num} $(b)$. We see that the $\mu_k$ with $0 \leq k
\leq \frac{N}{2}$ grow
regularly as $v$ grows until $v_c = \frac{1 + h^2}{2}$, 
while all the $\mu_k$ with $k >
\frac{N}{2}$ remain zero, so that also the corresponding Betti numbers
must vanish. But at $v_c$ a dramatic event occurs,
because for all the values of $v > v_c$ the Betti numbres $b_k$ {\em
must} be those of an $N$-torus, which are reported for comparison in
Fig.~\ref{fig_morse_num} $(c)$. A sudden transition from the situation
depicted in Fig.~\ref{fig_morse_num} $(b)$ to that of
Fig.~\ref{fig_morse_num} $(c)$ occurs at $v_c$, i.e., 
$\frac{N}{2}$ Betti numbers simultaneoulsy become nonzero.

Now we are in a position to explain how these topological transitions 
can be described by topological invariants. The
optimal situation would be the possibility of computing all the $N+1$ Betti
numbers of the manifolds $M_v$ as a function of $v$: unfortunately, we
are only able
to set an upper limit to them, using the Morse inequalities
(\ref{morse_ineq}). 
Nevertheless, we can use Eqs.\ (\ref{chi_morse}),
(\ref{n(v)}) and (\ref{mu_k})
to compute the Euler characteristic of the manifolds $M_v$, since  we only
need the $\mu_k$ for that. 
It turns out then that $\chi$ jumps from positive to negative
values, so that it is easier to look at $|\chi|$. 
In Figure \ref{fig_chi},  $\log(|\chi|(M_v))/N$ is plotted as a function of
$v$ for various values of $N$ ranging from 50 to 800. 
The ``big'' topology change occurring at the maximum value of $\cal V$, which
corresponds in the thermodynamic limit to the phase transition, implies a
discontinuous jump of $|\chi|$ from a big value to zero.

\section{The one-dimensional $XY$ model}
\label{sec_1d}

We consider now one example where there are topological changes very similar 
to the ones of the mean-field $XY$ model but no phase 
transition occurs, i.e., the one-dimensional XY model with nearest-neighbor 
interactions, whose Hamiltonian is of the class (\ref{ham_standard}) 
with interaction potential
\begin{equation}
V (\varphi) = \frac{1}{4}\sum_{i=1}^N \left[ 1 - 
\cos(\varphi_{i+1} - \varphi_i) \right] -h\sum_{i=1}^N \cos\varphi_i ~.
\label{V_nn}
\end{equation}
In this case the configuration space $M$ is still an $N$-torus, 
and using again the interaction energy per degree of freedom ${\cal V} = V/N$ 
as a Morse function, we can prove that also here 
there are many topological changes in the submanifolds 
$M_v$ as $v$ is varied from its minimum to its maximum value. 
 
The critical points are again those where the $\varphi_i$' s are equal either 
to $0$ or to $\pi$. However, at variance with the mean-field XY model, 
the critical values are now determined by the number of domain walls, $n_d$, 
i.e., the number of boundaries between connected regions on the chain where 
the angles are all equal (``islands'' of $\pi$'s and ``islands'' of $0$'s). 
The number of $\pi$'s leads only to a correction ${\cal O}(h)$ to the
critical value of $v$, which is given by 
\begin{equation}
v(n_d;n_\pi) = \frac{n_d}{2N} + h\,n_\pi~.
\label{v(nd)_1d}
\end{equation}
Since $n_d \in [0,N-1]$ (with free boundary conditions, $n_d =
0,1,\ldots,N-1$, 
while with periodic
boundary conditions $n_d$ is still bounded by $0$ and $N-1$,
 but can only be even), 
the critical values lie in the same interval as in the case of the mean-field 
XY model. But now the maximum critical value, 
instead of corresponding to a huge number of critical points, 
which rapidly grows with $N$, corresponds to {\em only two} 
configurations with $N-1$ domain walls, which are $\varphi_{2k} = 0$, 
$\varphi_{2k + 1} = \pi$, with $k = 1,\ldots,N/2$, and the reversed one.

\begin{figure}
\epsfysize= 14 truecm 
\begin{center}
\mbox{\epsfbox{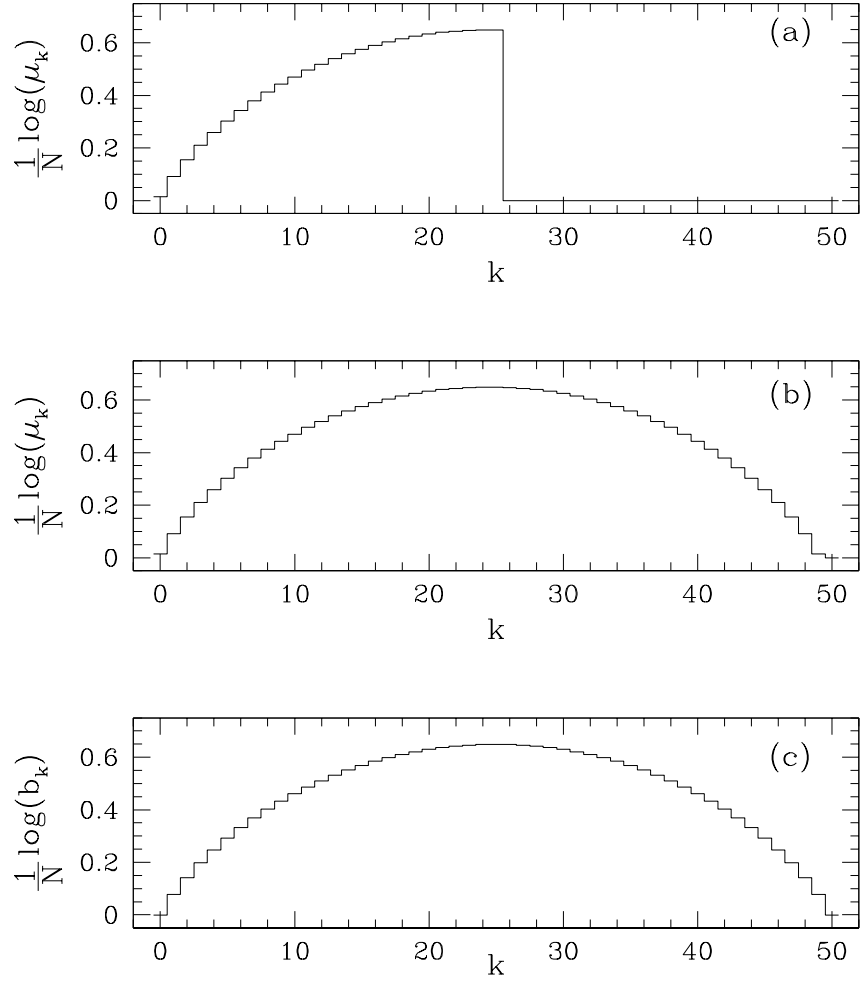}} 
\end{center}
\caption{The same as Fig.~\protect\ref{fig_morse_num} for the one-dimensional 
$XY$ model with nearest-neighbor interactions. 
$(a)$ Histogram of $\log(\mu_k(M_v))/N$ as a function of $k$ for
$v=1/4$; $(b)$ Histogram of $\log(\mu_k(M_v))/N$ as a function of $k$ for
$v=1/2$. In both cases $N =$
50 and $h = 0.01$.
$(c)$ For comparison, histogram of $\log(b_k(\protect\mathbb{T}^N))/N$ 
as a function of $k$ for
an $N$-torus $\protect\mathbb{T}^N$, with $N=50$.}
\label{fig_mu_1d}
\end{figure}

The number of critical points with $n_d$ domain walls is therefore 
(assuming free boundary conditions)
\begin{equation}
N(n_d) = 2{{N-1} \choose n_d}~.
\label{N_1d}
\end{equation}
We can compute the index of the critical points also in this case (see
Appendix \ref{app_proof_1d} for details). It turns out that 
\begin{equation}
{\rm index}(n_d) = n_d~,
\label{index_1d}
\end{equation}
so that 
\begin{equation}
\mu_k(n_d) = 2{{N-1} \choose k} \, \Theta(n_d - k)~.
\end{equation}
It is evident then that any topology change here corresponds to the 
attaching of handles of the same type. However, no ``big'' change 
like the one at $v_c$ in the case of
the mean-field model exists, although 
$\cal V$ is a Morse-Smale function on the 
whole manifold $M$. To illustrate this, we plot in Fig.~\ref{fig_mu_1d} 
the values
of the Morse indices $\mu_k$ as a function of $k$, as we have already done for
the mean-field $XY$ model in Fig.~\ref{fig_morse_num}. Comparing
Fig.~\ref{fig_mu_1d} with Fig.~\ref{fig_morse_num}, we see that while
in the mean-field model there is a critical value where $\frac{N}{2}$
Betti numbers become simultaneously nonzero, i.e., there exists a
topology change which corresponds to
the simultaneous attaching of handles of $\frac{N}{2}$ different
types, while here in the one-dimensional nearest-neighbor model nothing 
like that happens.

\begin{figure}
\epsfysize= 8 truecm 
\begin{center}
\mbox{\epsfbox{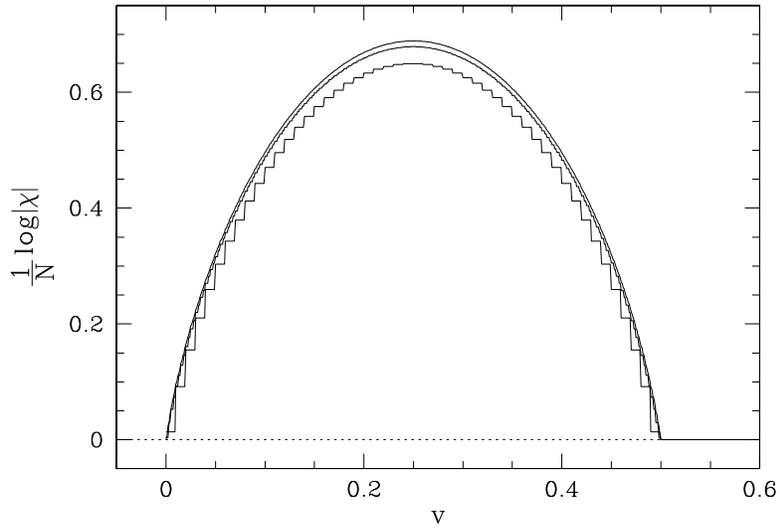}} 
\end{center}
\caption{Plot of $\log(|\chi|(M_v))/N$ for the one-dimensional $XY$ model with
nearest-neighbor interactions as a function of $v$. $N=$ 50, 200, 800
(from bottom to top).}
\label{fig_chi_1d}
\end{figure}

Also in this case, using Eqs.~(\ref{N_1d}) and (\ref{index_1d}), 
we can compute the Euler characteristic of the submanifolds $M_v$:
\begin{equation}
\chi(M_v) = 2\sum_{k=0}^{n_d(v)} (-1)^k {{N-1} \choose k} = 2\, (-1)^{n_d(v)} 
{{N-2} \choose n_d(v)}~,
\label{chi_1d}
\end{equation}
where, due to Eq.~(\ref{v(nd)_1d}),  
\begin{equation}
n_d (v) = 2Nv + {\cal O}(h)~.
\end{equation}
The Euler characteristic for the one-dimensional nearest-neighbor case is 
shown in Fig.~\ref{fig_chi_1d}. Comparing this figure with
Fig.~\ref{fig_chi}, we see that there is here no jump in the Euler
characteristic.

Before discussing the relevance that these results may have for the
general problem of the relation between topology and phase
transitions, it is illustrative to consider two abstract simplified models 
of topological transitions that occurred in the two models we have considered 
so far.

\section{Two-dimensional model of topology changes}
\label{sec_model}

\begin{figure}
\epsfysize= 6.5 truecm 
\begin{center}
\mbox{\epsfbox{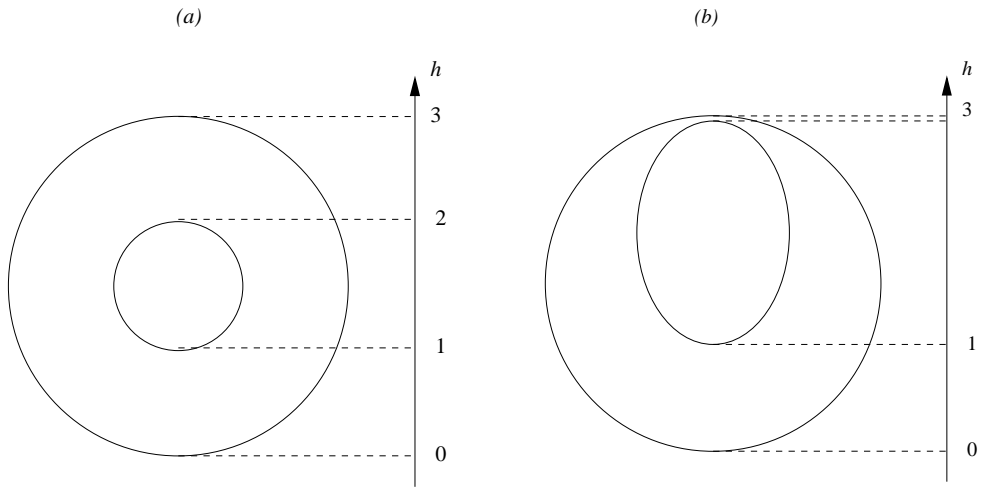}} 
\end{center}
\caption{$(a)$ A torus $\protect\mathbb{T}$ 
and its height function. Here $h_{\rm max} = 3$. 
$(b)$ A deformation $M$ of such
a torus as explained in the text.}
\label{fig_torus}
\end{figure}

A two-dimensional model of
the topological transition occurring in the configuration space of the physical
models we have discussed, which could perhaps help the intuition, can be
built as follows. Let us consider a two-dimensional torus $\mathbb{T}$, 
and place
it on a plane: let $h_{\rm max}$ be the maximum height of the surface above the
plane (see Fig.~\ref{fig_torus} $(a)$). 
Then deform the torus 
(by means of a diffeomorphism) until the upper end of the hole
is at height $h_{\rm max} - \varepsilon$, obtaining the surface $M$
shown in Fig.~\ref{fig_torus} $(b)$. 
It is apparent that $\varepsilon$ can be
made as small as we want. 

Let us now consider the height function $\cal H$ 
above the plane as a Morse function, so that the manifolds
\begin{eqnarray}
\mathbb{T}_h & = & {\cal H}^{-1} (-\infty,h] = 
\{ x \in \mathbb{T} : {\cal H}(x) \leq h\} \\
M_h & = & {\cal H}^{-1} (-\infty,h] = 
\{ x \in M : {\cal H}(x) \leq h\} 
\end{eqnarray}
are defined. As $h$ varies from its minimum to its maximum values ($h = 0$
and $h=3$, respectively, in Fig.~\ref{fig_torus}), the manifolds
$\mathbb{T}_h$ and $M_h$ cover the whole torus; as long as $h$ is lower
than the top of the hole ($h = 2$ in Fig.~\ref{fig_torus}), 
both  $\mathbb{T}_h$ and $M_h$ are ``half-tori'', but then 
the $\mathbb{T}_h$ become \emph{gradually} a full torus, 
while the $M_h$ jump abruptly from a half-torus to a
full torus as $h$ is changed by $\varepsilon$. Identifying the height function
with the potential energy, the case of the $\mathbb{T}_h$ clearly
reminds of the
behavior of the one-dimensional $XY$ model with no phase transition, while the
case of the $M_h$ seems close to what happens in the mean-field $XY$ model, and
the ``jump'' from the half-torus to the full torus is similar to the topology
change which is connected to the phase transition.

\begin{figure}
\epsfysize= 5 truecm 
\begin{center}
\mbox{\epsfbox{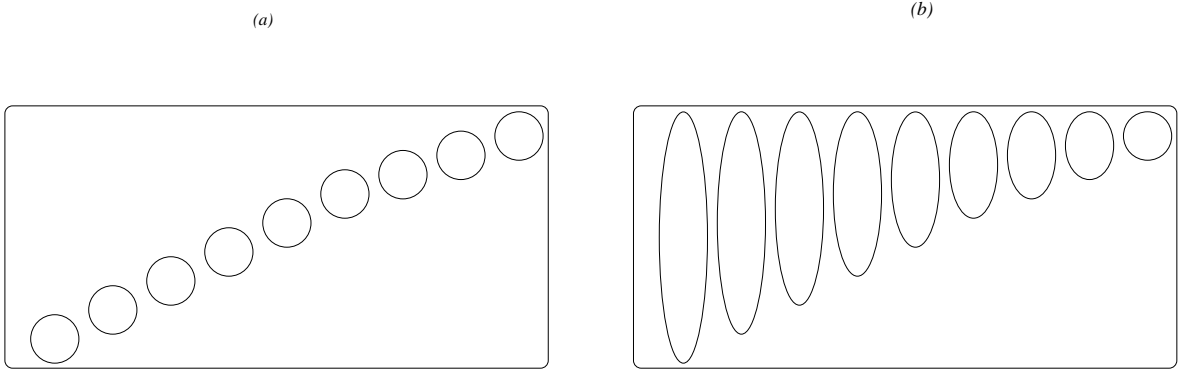}} 
\end{center}
\caption{$(a)$ A surface
of high genus (here $g=9$). $(b)$ A deformation $ {M}^{(g)}$ of such
a surface as explained in the text.}
\label{fig_surface}
\end{figure}

The analogy with the mean-field models 
becomes even clearer if, instead of a torus, we consider a compact 
surface of genus\footnote{The {\it genus} $g$ is the number of handles of a 
two-dimensional surface.} $g \gg 1$, i.e., with many holes, and deform
the surface with a diffeomorphism until the upper end of all the holes is at
height $h_{\rm max} - \varepsilon$, as shown in Fig.~\ref{fig_surface}. 
Again, $\varepsilon$ can be
made as small as we want. Let us denote by ${M}^{(g)}$ the deformed
surface.

\begin{figure}
\epsfysize= 8 truecm 
\begin{center}
\mbox{\epsfbox{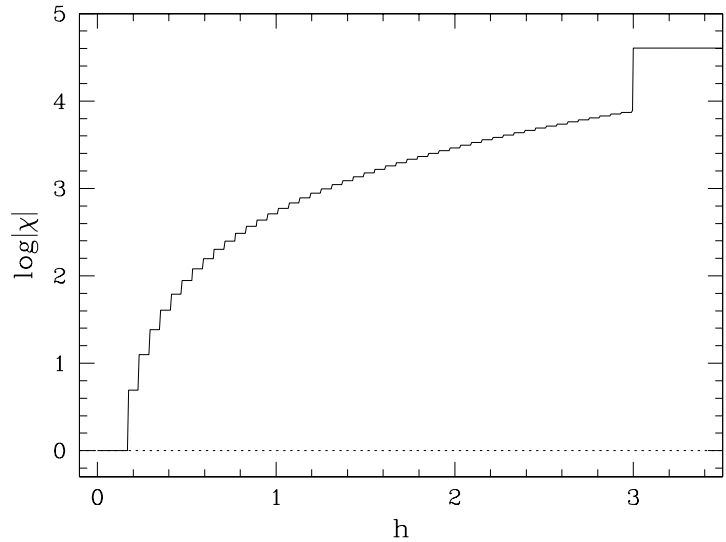}} 
\end{center}
\caption{Plot of the logarithm of the absolute value of the 
Euler characteristic of the submanifolds
of given height of a surface $ {M}^{(g)}$ like the one depicted 
in Fig.~\protect\ref{fig_surface} $(b)$, with $g = 50$ and $h_{\rm max} = 3$, 
as a function of the height $h$. The small jumps are the topology
transitions corresponding to the crossing of the bottoms of the holes: the 
last big jump is the one occurring at $h_{\rm max}$, when $\varepsilon \to 0$.}
\label{fig_chi_model}
\end{figure}

The Betti numbers of the surface $ {M}^{(g)}$ are \cite{spivak}
\begin{equation}
b_0 = b_2 = 1\,, \qquad b_1 = 2g\,,
\end{equation}
and the Euler characteristic is
\begin{equation}
\chi(M^{(g)}) = \chi( {M}^{(g)}) = 2 - 2g\, ,
\end{equation}
i.e., a big negative number. Let us now consider, 
as in the previous case of the torus, the height function $\cal H$ 
above the plane as a Morse function. The manifolds
\begin{equation}
 {M}^{(g)}_h = {\cal H}^{-1} (-\infty,h] = 
\{ x \in  {M}^{(g)} : {\cal H}(x) \leq h\}
\end{equation}
will be topologically very different from the whole $M^{(g)}$ as long as 
$h < h_{\rm max} - \varepsilon$ but sufficiently large that all the critical
levels corresponding to the bottoms of all the holes have already been
crossed: in fact, their Betti numbers will be
\begin{equation}
b_0( {M}^{(g)}_h) = 1,~b_1( {M}^{(g)}_h) = g,~b_2( {M}^{(g)}_h) 
= 0\,,
\end{equation}
and the Euler characteristic will be
\begin{equation}
\chi( {M}^{(g)}_h) = 1 - g\,.
\end{equation}
Then, by changing the value of the height by an amount $\varepsilon$ as small 
as one wants, one changes the Betti number
$b_1$ from $g$ to $2g$, the $b_2$ from $0$ to $1$ and $\chi$ from $1 - g$ to
$2-2g$. This is a topological change which involves a change of ${\cal O}(d)$
Betti numbers ($d$ is the dimension of the manifold); moreover, the size of the
change is of the order of the value of the Betti numbers. This topology change
involves also a change
of the Euler characteristic $\chi$ which is again of the same order as its 
value. Identifying again the height function with the potential energy, we see
that this is just what happens in the case of the
$M_v$ of the mean-field $XY$ model, although there the dimension of the 
manifolds is $N$ and very large, while in this low-dimensional analogy it is
only $d=2$. The behavior of $|\chi( {M}^{(g)}_h)|$ 
as a function of $h$ is plotted in
Fig.~\ref{fig_chi_model}. We see that the behavior of $|\chi|$ is indeed very
similar to the case of the mean-field $XY$ model, the only big difference being
that in the latter $|\chi|$ jumps to zero while here it jumps to a nonzero
value. However this difference is due to the fact that the Euler
characteristic of a torus is zero while that of a surface of genus $g$
is $2-2g$.
 
\section{Concluding remarks and open questions}
\label{sec_comments}

We conclude with some comments and remarks, and with a discussion of the open
problems and of the future perspectives opened by the results we have 
presented.

\subsection{Topology and thermodynamic functions}
\label{sec_comm_topotd}
The results we have reported in the present paper concerning the mean-field and
the one-dimensional $XY$ models, together with a 
theorem \cite{theorem} which states the necessity of topology changes in the
$M_v$'s for the existence of phase transitions (at least for a certain class of
short-range interactions), lend strong support to the TH. However, one could
wonder whether a direct relation between topology and thermodynamic quantities,
like entropy or temperature, exists.

A result -- although approximate -- which shows that there could be an explicit
contribution of topological quantities to the temperature was indeed found in a
previous paper \cite{CSCP}: it relates the inverse temperature
$\beta = \frac{1}{T}$ to the Betti numbers $b_i$ of the constant-energy 
hypersurface
in phase space, $\Sigma_E$. We stress the fact that such a relation -- Eq.~(43)
in Ref.~\cite{CSCP} -- does not allow to express $\beta$ in terms of 
topological
invariants alone, because also other contributions, of a non-topological 
nature,
exist: nonetheless a topological contribution expressed as a function of the 
Betti numbers of $\Sigma_E$ is explicitly present. A similar relation can be 
obtained also for the entropy: the derivation only involves a slight 
modification 
of the reasoning of Sec.~V of Ref.~\cite{CSCP}. Without entering into
technical details, we simply observe that, under the same approximations 
as made in Ref.~\cite{CSCP}, the ``topological contribution'' $\tau(E)$ to the 
entropy per degree of freedom $s = \frac{S}{N}$ can be written, up to constant 
terms, as 
\begin{equation}
\tau(E) \simeq \frac{1}{2N-1} \log \sum_{i=0}^{2N - 1} b_i(\Sigma_E)~,
\label{tau(E)}
\end{equation}  
an expression identical to the corresponding expression for $\beta$ 
in \cite{CSCP}, apart from the presence of a logarithm here. 
We remark that, as in the case of $\beta$, there are also other contributions, 
of a
non-topological nature, to the entropy. However, it is interesting to look at 
this topological $\tau$-contribution alone to see whether it carries the 
relevant information about
the way the topological transition of the energy hypersurface in phase space 
triggers
the phase transition, i.e., the nonanalyticity of the thermodynamic quantities.
Therefore we can try to compute $\tau$ for the models studied in the present 
paper.

First of all, we have to express $\tau(E)$ in terms of the topological 
invariants of the manifolds $M_v$ instead of in terms of the Betti numbers 
of the energy hypersurfaces $\Sigma_E$ [Eq.(\ref{tau(E)})].
To do that, we resort to the fact that -- at large $N$ and for systems with 
standard 
kinetic energy $K$ -- the volume measure of $\Sigma_E$ concentrates 
on\footnote{$\mathbb{S}^{N-1}_{\langle 2K \rangle^{1/2}}$ is an 
$N-1$-dimensional hypersphere defined by the kinetic energy term
$\sum p_i^2=2\langle K\rangle$. We have $E=\langle K\rangle +\delta K +
\langle V\rangle +\delta V$, and $\delta K/\langle K\rangle,
\delta V/\langle V\rangle ={\cal O}(1/\sqrt{N})$. } 
$\mathbb{S}^{N-1}_{\langle 2K 
\rangle^{1/2}}\times M_{\langle
V\rangle}$, so that $\Sigma_E$ can be approximated by this product manifold.
Using then the Kunneth formula \cite{nakahara} for the Betti numbers of
a product manifold $A\times B$, i.e.,
\begin{equation}
b_i(A\times B) = \sum_{j+k=i} b_j(A)\, b_k(B)
\end{equation}
which, when applied to 
$\mathbb{S}^{N-1}_{\langle 2K 
\rangle^{1/2}}\times M_{\langle
V\rangle}$, gives $b_i(\Sigma_E)= 2 b_i(M_v)$ for $i = 1,\ldots,N-1$,
and $b_j(\Sigma_E)= b_j(M_v)$ for $j = 0,N$, since all the Betti numbers of a 
hypersphere vanish except $b_0$ and $b_N$ which are equal to 1. 
This allows us to write
\begin{equation}
\tau(v)\simeq \frac{1}{2N}\log \left[ b_0(M_v) +
2\sum_{i=1}^{N-1}
b_i(M_v ) + b_N(M_v) \right] ~,
\label{tau(v)}
\end{equation}
where we have also replaced $2N - 1$ with $2N$ because we assume that $N$ is 
large.
Since we cannot compute
the Betti numbers exactly, we cannot evaluate this expression;
nonetheless, we can estimate it by using the Morse numbers as
approximations of the Betti numbers, 
i.e., by putting
\begin{equation}
b_i(M_v) \simeq \mu_i(M_v)~, \qquad i=0,\ldots,N~.
\end{equation}
In general, with the exception of perfect Morse functions (i.e. those for 
which $b_i=\mu_i$),
the Morse numbers are only upper bounds of the Betti numbers. However, 
looking at
Figs.~\ref{fig_morse_num} (b)-(c) (for the mean-field $XY$ model) and 
\ref{fig_mu_1d}
(b)-(c) (for the one-dimensional $XY$ model), we can observe that the 
nonvanishing
Morse numbers of the manifolds $M_v$ for values of $v$ very close to 
$v_{\rm max}$
are very close to the Betti numbers of the $M_v$'s for $v = v_{\rm max}$
(which are $N$-tori), and this suggests that for the models studied in 
the present
paper the Morse numbers $\mu_i$ could well be good approximations to the Betti
numbers $b_i$. 

\begin{figure}
\epsfysize= 8 truecm 
\begin{center}
\mbox{\epsfbox{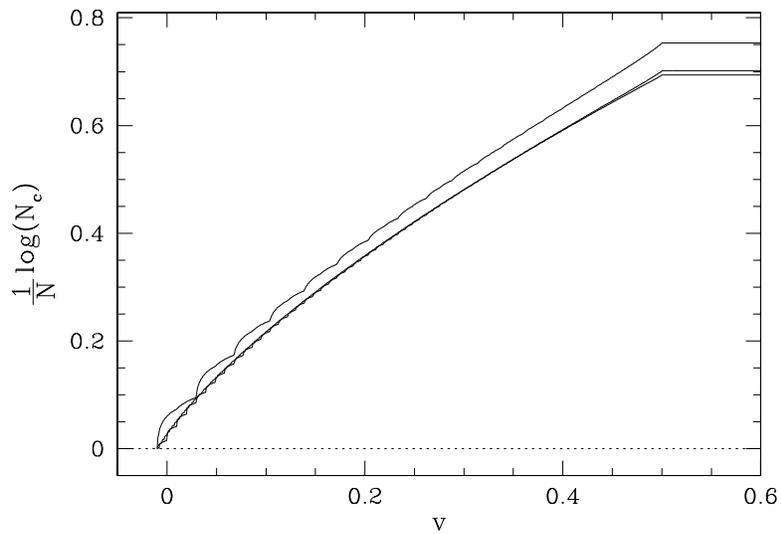}} 
\end{center}
\caption{Mean-field $XY$ model. 
Plot of $\log(N_c)/N$ as a function of $v$. $N =$
50,200,800 (from top to bottom) and $h = 0.01$; $v_c = 0.5 + {\cal O}(h^2)$.}
\label{fig_ncp}
\end{figure}

\begin{figure}
\epsfysize= 8 truecm 
\begin{center}
\mbox{\epsfbox{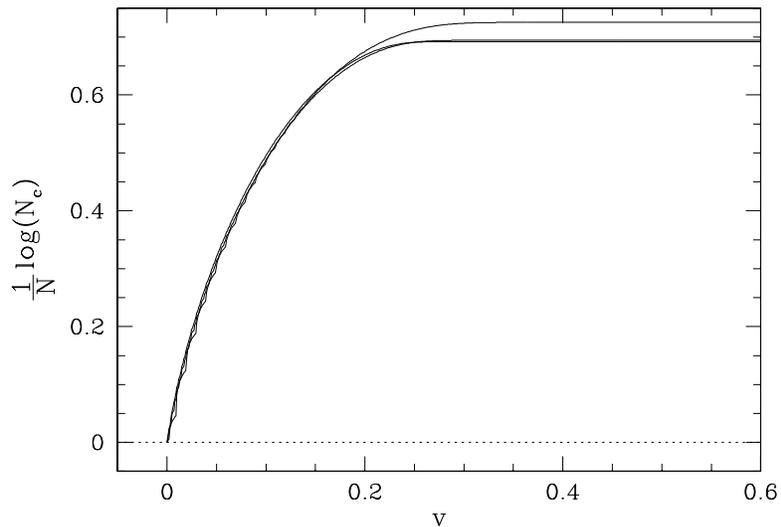}} 
\end{center}
\caption{One-dimensional $XY$ model with nearest-neighbor interactions.
Plot of $\log(N_c)/N$ as a function of $v$. $N =$
50,200,800 (from top to bottom) and $h = 0.01$.}
\label{fig_ncp1d}
\end{figure}

We can then compute the quantity
\begin{equation}
\tilde\tau (v) = \frac{1}{2N} \log \left[\mu_0(M_v) + 2
\sum_{i = 1}^{N-1} \mu_i(M_v)
+ \mu_N(M_v) \right] ~,
\end{equation}
for the two models studied in the paper. We note that 
\begin{equation}
\sum_{i = 0}^{N} \mu_i(M_v) = N_c(M_v)~,
\end{equation}
where $N_c(M_v)$ is the total number of critical points of the
function $\cal V$ in the manifold $M_v$. Therefore, 
when $N$ gets large, and if $\mu_0$ and $\mu_N$ are not much
larger than the other Betti numbers (which is true in our case, see
Fig.~\ref{fig_morse_num}), we can write approximately
\begin{equation}
\tilde\tau (v) \simeq \frac{1}{N} \log N_c(v)~,
\label{tau_approx}
\end{equation}
apart from an additive constant. The behavior of this quantity is plotted as a
function of $v$ in Fig.~\ref{fig_ncp} for the mean-field $XY$ model and in
Fig.~\ref{fig_ncp1d} for the one-dimensional $XY$ model.
First, we note that in both cases the ``topological contribution'' $\tilde\tau$
to the entropy behaves qualitatively as expected for the configurational 
entropy,
i.e., it grows monotonically up to the maximum value of $v$, after which it
remains constant. Moreover, we see
that, in the case of the mean-field model (Fig.~\ref{fig_ncp}), the topology 
change
at $v_c$ related to the phase transition 
corresponds to a discontinuity in the slope of $\tilde\tau(v)$, which thus 
seems to be both the precursor and
the source of the nonanalyticity of the entropy at the phase
transition point as $N\rightarrow\infty$. In the case of the one-dimensional 
$XY$ model, where no phase transition is present, unlike the
mean-field case, the curve is smooth for any $v$, consistent with
the fact that also the entropy of the system is smooth.

Therefore the contribution of the Betti numbers alone to the thermodynamic 
quantity
$s$ appears to yield the phase transition point correctly; an indication 
again of the topological mechanism of the phase transition.

\begin{figure}
\epsfysize= 8 truecm 
\begin{center}
\mbox{\epsfbox{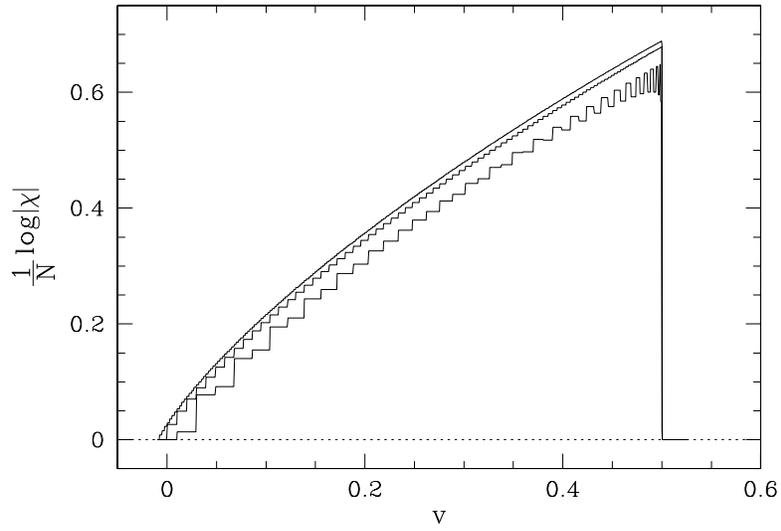}} 
\end{center}
\caption{Mean-field $XY$ model. 
Plot of $\log(|\chi|(M_v))/N$ as a function of $v$. $N =$
50,200,800 (from bottom to top) and $h = 0.01$; $v_c = 0.5 + {\cal O}(h^2)$.}
\label{fig_chi}
\end{figure}

We emphasize that the discontinuity in Fig.~\ref{fig_chi} at $v = v_c$ 
corresponds
to that of Fig.~\ref{fig_ncp} at the same value of $v$. Even more, the many 
small
jumps in the Euler characteristic occurring in Fig.~\ref{fig_chi}, and  
corresponding to
the topological changes which occur at $v < v_c$, are smoothed out in 
Fig.~\ref{fig_ncp}
where the topological contribution to the entropy, $\tilde\tau$, is reported. 
This is
due to the fact that while the Euler characteristic is the alternating sum of 
the Morse
numbers $\mu_i$, the $\tau$ is the sum of them: this is a further indication 
that to
yield a phase transition, i.e., a discontinuity in a derivative of $s$, a 
``strong''
topology change would be needed, such as to affect the variation of the sum 
of all the Betti numbers as a function of $v$. 

However, to what extent these results are of general validity remains an 
interesting
open question.

\subsection{When does a topology change entail a phase transition?}

Another fundamental point which still remains open in the topological approach 
to phase
transitions is the question of which are the {\em sufficient} conditions for
the topology changes of the manifolds $M_v$ to entail a phase transition.
Topology changes appear to be rather common, and most of them are {\em not}
connected to phase transitions.  

In our previous papers, clear evidence -- albeit only numeric -- was 
found that phase transitions would correspond to very significant
transitions in the way the topology changes \cite{phi4} as a function of $v$. 
Moreover, in Sec.~\ref{sec_comm_topotd} we have discussed a possible
mechanism through which a topological change could trigger a change in the 
properties
of thermodynamic functions, resulting in a phase transition.

On the basis of our results obtained before and here, we  put forward the
conjecture that what we have observed in the case of $XY$ models may well have
general validity: a topology change in the submanifolds $M_v$ might entail a
phase transition if it involves the simultaneous attachment of handles of 
${\cal O}(N)$ different types on the same critical level. 
However, we must be aware that this could well be specific to the class of 
models studied here: to what extent this conjecture might have general 
validity remains an open question.

\subsection{The role of the external field $h$}
 
Studying the topology changes in the configuration space of $XY$ models, 
mean-field as well as one-dimensional,  we have considered the presence of an
external field $h \not= 0$  which explicitly breaks the $O(2)$-invariance of
the potential energy, and then, discussing the connection with phase
transitions,  we have considered the case in which $h$ tends to zero. We did
that for the sake of simplicity, for, if we set $h = 0$ from the outset, the
potential energy per degree of freedom $\cal V$ is {\em not}, rigorously
speaking, a Morse function, because its $O(2)$-invariance entails the presence
of a zero eigenvalue in its Hessian. When $h=0$, the critical points of $\cal
V$ are {\em not} isolated, but form one-dimensional manifolds (topologically
equivalent to circles) which are left unchanged by the action of the $O(2)$
continuous symmetry group so that the critical points become in this case 
critical manifolds.
However, in the case of the mean-field $XY$ model, as far as the presence and 
the nature of topology changes are concerned, studying
the case with $h=0$ from the outset we find {\em exactly} the same behavior as
in the case we have discussed in this paper, i.e., as long as $v < v_c$ only
handles of the same type are attached, while at $v = v_c$ handles of $N/2$
different types are attached, the only difference between the two cases being 
that when $h=0$ the handles are not attached at isolated points, but rather 
to the entire critical manifold \cite{milnor,Palais}.
However, putting $h = 0$ from the beginning makes 
the computation of the Euler characteristic $\chi(M_v)$ via the Morse
numbers much more difficult, 
because now one has to take into account the contributions to $\chi$
coming from the Betti numbers of
the critical manifolds  (see Ref.~\cite{Dubrovin} for the details).

In the case of the one-dimensional nearest-neighbor $XY$ model, where no 
phase transition is
present, the use of $h=0$ from the outset implies a further complication, i.e.,
that the critical
points consist not only of 
the configurations made of $0$'s and $\pi$'s, but also of
spin waves, that is configurations 
\begin{equation}
\varphi_j = \varphi_0 \, e^{i k j}~,
\end{equation} 
with wavenumbers $k$ depending on boundary conditions.

For all these reasons we preferred to force the potential energy to be a Morse
function via the explicit breaking of the $O(2)$ symmetry using an external
field $h\neq 0$. Incidentally, we notice that in the mean-field
$XY$ model, as long as $h \not = 0$, the topology changes which do not
correspond to any phase transition (i.e, those occurring at $v < v_c$) occur at
a number of values of $v$ which grows with $N$, and these values 
become closer and closer
as $N$ grows, eventually filling the whole interval $[0,\frac{1}{2}]$ as $N \to
\infty$. To the contrary, the value $v_c$, which corresponds to the ``big'' 
topology
change connected to the phase transition, remains separated from the others by 
an
amount ${\cal O}(h^2)$ also in the thermodynamic limit, and tends to
$\frac{1}{2}$ only when $h \to 0$. This is reminiscent
of a similar fact occurring
in statistical mechanics, where one observes a spontaneous symmetry breaking,
signalled e.g.\ by the onset of a finite magnetization even at zero external
field, if one assumes the
presence of an external field and then lets it tend to zero only after the
thermodynamic limit is taken. 
\subsection{Transitional phenomena in finite systems and other future
developments}

$(a)$ 
If it would be possible to establish a one-to-one correspondence between a
particular class of topology changes in configuration space and the usual
thermodynamic phase transitions defined in the thermodynamic limit, then, as a
by-product, one would have available also a natural definition of phase
transitions for {\em finite} systems. In fact, the topology changes are defined
at any $N$, so that one would call ``phase transitions at finite $N$'' those
topology changes in configuration space which become phase transitions in the
usual, statistical-mechanical sense at infinite $N$. This would be an
interesting consequence 
of the present topological approach to phase transitions, since it would
circumvent  the basic problem of any clear 
definition of phase transitions at finite
$N$, by showing that all phase transitions have their basic origin in
a topological change which may occur also at finite $N$, but entail true
mathematical  singularities in the thermodynamic functions only at infinite
$N$.

$(b)$ 
Finite systems do not seem to be the only active field of statistical physics
where the topological approach might prove useful. Another promising field of
application is that of glasses or, in general, of disordered systems. It is now
quite clear that many of the puzzling properties of glasses are encoded in
their ``energy landscape'' \cite{glasses}, i.e., in the structure of valleys
and saddles of the potential energy function: but this is directly connected to
the structure of the submanifolds $M_v$ and $\Sigma_v$ of configuration space,
and in fact topological concepts start to emerge in some recent papers on
glasses \cite{Angelani}.

$(c)$ 
Finally, we notice that at present only systems undergoing second-order phase
transitions have been studied via the topological approach: a natural
question which arises is then if also first-order transitions can be explained
topologically. Work is in progress in this direction, and some preliminary
results indicate that also discontinuous phase transitions can be connected to
topology changes of the submanifolds $M_v$, and that a signature of them can 
be found in their Euler characteristic.

\begin{acknowledgments} 
We warmly thank R.~Franzosi, L.~Spinelli, and 
G.~Vezzosi for very helpful discussions and suggestions.  This work 
is part of the INFM PAIS
{\em Equilibrium and Non-Equilibrium Dynamics of condensed matter
systems}.
EGDC gratefully acknowledges support from the Office of Basic
Engineering Science of the US Department of Energy, under Grant
No.~DE-FG 02-88-ER 13847, and from the IS-Sabbatical of the INFM. 
\end{acknowledgments}

\appendix

\section{Estimate of the index of the critical points}
\label{app_proof}

In this Appendix we want to discuss some details related to the
estimates of the indices of the critical points we have used in
Sec.~\ref{sec_MF} and Sec.~\ref{sec_1d}. 

\subsection{Mean-field $XY$ model}
\label{app_proof_MF}
  
Here we want to prove the crucial estimate (\ref{estimate}), which
 we used in Sec.~\ref{sec_MF} to compute the
index of the critical points for the mean-field $XY$ model.

We recall that we want to compute the the number of negative eigenvalues
of the Hessian matrix of the function $\cal V$, i.e., of the matrix $H$ whose
elements $H_{ij}$ are
\begin{equation}
H_{ij} = \frac{\partial^2 {\cal V}}{\partial \varphi_i \partial \varphi_j}  
\qquad i,j = 1,\ldots, N\,,
\end{equation} 
where ${\cal V} = V/N$ and $V$ is the potential energy of the mean-field $XY$
model defined in Eq.~(\ref{V}). 
The diagonal elements of this matrix are
\begin{equation}
H_{ii} = d_i = \frac{1}{N}\left[(m_x + h) \cos\varphi_i + m_y
\sin\varphi_i\right] - \frac{1}{N^2}~,
\end{equation}
and the off-diagonal ones are
\begin{equation}
H_{ij} = - \frac{1}{N^2}\left(\sin\varphi_i\sin\varphi_j +
\cos\varphi_i\cos\varphi_j \right)\,.
\label{nondiag_app}
\end{equation}
At the critical points of $\cal V$, the angles are either $0$ or $\pi$, so that
the sines are all zero and the cosines are $\pm 1$. Moreover, since we are
interested only in the signs of the eigenvalues of $H$ and not in their absolute
values, we multiply $H$ by $N$ in order to get rid of the $1/N$ factor in front
of it. We can then write the matrix $H$ (multiplied by $N$) as
\begin{equation}
H = D + B
\label{matrixH}
\end{equation}
where $D$ is a diagonal matrix,
\begin{equation}
D = {\rm diag}(\delta_i)\,,
\end{equation}
whose elements $\delta$ are 
\begin{equation}
\delta_i = (m_x + h) \cos\varphi_i ~,
\end{equation}
where the $\varphi_i$'s ($i = 1,\ldots,N$) are computed at the
critical point, and the elements of $B$ can be written in terms of a vector
$\sigma$ whose $N$ elements are either $1$ or $-1$:
\begin{equation}
b_{ij} = - \frac{1}{N} \sigma_i \sigma_j\,,
\label{bij}
\end{equation}
where
\begin{equation}
\sigma_i = +1 (-1) ~{\rm if}~ \varphi_i = 0 (\pi)\,.
\end{equation}
This, since when the angles are either $0$ or $\pi$, the sines 
in Eq.\ (\ref{nondiag}) vanish, so that then
\begin{equation}
N H_{ij} = - \frac{1}{N}\cos\varphi_i\cos\varphi_j\,,
\end{equation}
and 
\begin{equation}
\cos\varphi_i = \sigma_i \,.
\end{equation}

Having fixed the notation, 
our goal is to show that, at least when $N$ is large, the number of negative
eigenvalues of the full matrix $H$, i.e., the index of the critical point, 
can be conveniently approximated by the number of negative eigenvalues of $D$,
that is, by the number of negative $\delta$'s. To do that, we proceed in two
steps: $(i)$ we show that the matrix $B$ is of rank one (which implies that $B$
has $N-1$ zero eigenvalues and only one nonzero eigenvalue), 
and $(ii)$ we adapt a
theorem due to Wilkinson \cite{Wilkinson} 
to this case, thus proving our assertion.

As to step $(i)$, let us consider for example a case with $N=3$, and the
critical point corresponding to, say, $(\varphi_1,\varphi_2,\varphi_3) =
(\pi,0,0)$. The vector $\sigma$ is then
\begin{equation}
(\sigma_1,\sigma_2,\sigma_3) = (-1, 1, 1)~.
\end{equation} 
Using Eq.\ (\ref{bij}), the matrix $B$ is
\begin{equation}
- \frac{1}{3} \left( 
\begin{array}{ccc}
1 & -1 & -1 \\
-1 & 1 & 1 \\
-1 & 1 & 1
\end{array}
\right)~.
\end{equation}
We see that the second row is equal to the first multiplied by $-1$, and the
same holds for the third row. This is true for any $N$, and is a consequence of
Eq.\ (\ref{bij}): any row of the matrix $B$ is equal to another row multiplied
by either $+1$ or $-1$. This means that $N-1$ rows are not linearly independent
and that the rank of the matrix is one.

We have then proved that our Hessian matrix $H$ is the sum of a diagonal matrix
and of a matrix of rank one.

Let us now pass to step $(ii)$. First, we recall a theorem of
Wilkinson found in Ref.~
\cite{Wilkinson}: 

\begin{theorem}[Wilkinson]
Let $A$ and $B$ be $N\times N$ real symmetric matrices and let
\[
C = A + B~.
\]
Let $\gamma_i$, $\alpha_i$ and $\beta_i$ ($i = 1,2,\ldots,N$) be the (real)
eigenvalues of $C$, $A$ and $B$, respectively, arranged in non-increasing order,
i.e., 
\[
\begin{array}{ccccccc}
\gamma_1 & \geq & \gamma_2 & \geq & \ldots & \geq & \gamma_N\, ; \\ 
\alpha_1 & \geq & \alpha_2 & \geq & \ldots & \geq & \alpha_N\, ; \\ 
\beta_1 & \geq & \beta_2 & \geq & \ldots & \geq & \beta_N\, . 
\end{array}
\] 
Then
\begin{equation}
\gamma_{r + s -1} \leq \alpha_r + \beta_s \qquad \forall\, r+s -1 \leq N\, .
\label{W1}
\end{equation}
\end{theorem}
Notice that we can also write 
\begin{equation}
A = C + (-B)\, ,
\end{equation}
and since the eigenvalues of $-B$ arranged in non-increasing order are
\[
-\beta_{N - i + 1}
\]
we can write
\begin{equation}
\alpha_{r + s -1} \leq \gamma_r - \beta_{N - s +1}\, .
\label{W2}
\end{equation}

Now, we are interested in a special case, i.e., the case in which the matrix $B$
is of rank one (has only one nonzero eigenvalue). What does Wilkinson's theorem
say when applied to such a special case? We consider the two possible cases,
namely:
\begin{description}
\item[$(a)$] the nonzero eigenvalue is negative:
\[
\beta_i = 0 ~{\rm for}~ i = 1,2,\ldots,N-1, \beta_N = -\varrho\,;
\]
\item[$(b)$] the nonzero eigenvalue is positive:
\[
\beta_N = \varrho, ~\beta_i = 0 ~{\rm for}~ i = 2,3,\ldots,N\,.
\]
\end{description}

\noindent \underline{Case $(a)$}. Choosing $s = 1$ in Eq.\ (\ref{W1}) we get
$\beta_s = 0$, so that
\begin{equation}
\gamma_r \leq \alpha_r \qquad r = 1,\ldots,N\,,
\label{a1}
\end{equation}
while choosing $s = 2$ in Eq.\ (\ref{W2}) we get $-\beta{N-s+1} =0$ again, whence
\begin{equation}
\alpha_{r+1} \leq \gamma_r \qquad r = 1,\ldots,N -1\,.
\label{a2}
\end{equation}
Combining Eqs.\ (\ref{a1}) and (\ref{a2}) we obtain
\begin{equation}
\alpha_{r+1} \leq \gamma_r \leq \alpha_r \qquad r = 1,\ldots,N -1\,;
\label{at1}
\end{equation}  
\begin{equation}
\gamma_N \leq \alpha_N\,.
\label{at2}
\end{equation}  
We have thus shown that all the eigenvalues of $C$ (except for the smallest one)
are bounded between two successive eigenvalues of $A$. As to $\gamma_N$, we can
only say that it is smaller than (or equal to) the smallest eigenvalue of $A$. 

\medskip

\noindent \underline{Case $(b)$}. Choosing $s = 2$ in Eq.\ (\ref{W1}) we get
$\beta_s = 0$, so that
\begin{equation}
\gamma_{r+1} \leq \alpha_r \qquad r = 1,\ldots,N-1\,,
\label{b1}
\end{equation}
while choosing $s = 1$ in Eq.\ (\ref{W2}) we obtain
\begin{equation}
\alpha_r \leq \gamma_r \qquad r = 1,\ldots,N\,.
\label{b2}
\end{equation}
Combining Eqs.\ (\ref{b1}) and (\ref{b2}) we obtain
\begin{equation}
\alpha_{r} \leq \gamma_r \leq \alpha_{r-1} \qquad r = 2,\ldots,N\,;
\end{equation}  
\begin{equation}
\gamma_1 \geq \alpha_1\,.
\end{equation}  
We have thus shown again that all the eigenvalues of $C$ (except, in this case, 
for the largest one)
are bounded between two successive eigenvalues of $A$. As to $\gamma_1$, we can
only say that it is larger than (or equal to) the smallest eigenvalue of $A$,
$\alpha_1$.

Let us now apply these results to our problem, i.e., to the computation of the
number of negative eigenvalues of the matrix $H$ in Eq.\ (\ref{matrixH}). 
Denoting by $\eta_i$ its
eigenvalues, by $\delta_i$ those of $D$ and by $\beta_i$ those of $B$, we have
\begin{equation}
\beta_i = 0 \qquad \forall\, i \not = 1,N\,,
\end{equation}
and either 
\[
\beta_1 = 0, \beta_N = -\varrho\,,
\]
or
\[
\beta_1 = \varrho, \beta_N = 0\,.
\]
At a given critical point, with $n_\pi$ angles equal to $\pi$, 
the eigenvalues of $D$ are (for the moment we do not order them)
\begin{equation}
\delta_i = m_x + h \qquad i = 1,\ldots,N - n_\pi\,;
\end{equation}
\begin{equation}
\delta_i = -(m_x + h) \qquad i = N - n_\pi + 1,\ldots,N\,.
\end{equation}
The $x$-component of the magnetization vector is given by
\begin{equation}
m_x = 1 - \frac{2n_\pi}{N}
\end{equation}
so that 
\begin{eqnarray}
m_x & > & 0 \qquad {\rm if}~ n_\pi \leq \frac{N}{2}~, \\
m_x & < & 0 \qquad {\rm if}~ n_\pi > \frac{N}{2}~.
\end{eqnarray}
Then, if the external field $h$ is sufficiently small:\\
if $n_\pi \leq N/2$, then
\begin{equation}
\delta_i = m_x + h >0  \qquad i = 1,\ldots,N - n_\pi\,,
\end{equation}
\begin{equation}
\delta_i = -(m_x + h) <0 \qquad i = N - n_\pi + 1,\ldots,N\,,
\end{equation}
i.e., there are $N - n_\pi$ positive and $n_\pi$ negative $\delta$'s;\\
else if $n_\pi \leq N/2$, then
\begin{equation}
\delta_i = - (m_x + h) >0  \qquad i = 1,\ldots,n_\pi\,,
\end{equation}
\begin{equation}
\delta_i = m_x + h <0 \qquad i = n_\pi + 1,\ldots,N\,,
\end{equation}
i.e., there are $n_\pi$ positive and $N - n_\pi$ negative $\delta$'s.

Now we claim that, at least  as $N$ gets large, 
we can estimate the number of negative $\eta$'s, i.e., the
index of the critrical point, by saying that it is equal to the number of
negative $\delta's$. More precisely, we claim that 
the error of our estimate is not larger than 1,
i.e.,
\begin{equation}
{\rm index}(H) = \# (\eta < 0) = \# (\delta < 0) \pm 1~.
\label{estim}
\end{equation}
and as $N$ gets large this error becomes obviously negligible. 
To prove this statement,
let us consider the case in which $n_\pi < N/2$.
We observe that we do not know whether we are in case $(a)$ or in case $(b)$,
i.e., we do not know if the matrix $B$ has a negative or a positive eigenvalue.
But we can, anyway, 
try one of the two cases, say $(a)$. Using Eqs.\ (\ref{at1}) and
(\ref{at2}) we can then say that 
\begin{equation}
\delta_{r+1} \leq \eta_r \leq \delta_r < 0 
\qquad r = N - n_\pi + 1,\ldots,N - 1\,,
\end{equation}
(note that these are $n_\pi -1$ equations), and that
\begin{equation}
\eta_N \leq \delta_N < 0 \,.
\end{equation}
Thus we conclude 
that the number of negative $\eta$'s is just equal to that of negative
$\delta$'s, i.e., $n_\pi$. 
If we guessed correctly the sign of the nonzero eigenvalue of $B$,
then, our estimate is exact. But in case we guessed it wrong, i.e., if
we were in case $(b)$ and not $(a)$, by using Eqs.\  (\ref{at1}) and
(\ref{at2}), we would
have overestimated the number of negative $\eta$'s by 1. Conversely, if we had
used the equations of case $(b)$ in a situation which belonged to case $(a)$ we
would have underestimated the index by 1. So, 
we conclude that the error of our estimate is
always $\pm 1$.

\subsection{One-dimensional $XY$ model}
\label{app_proof_1d}

Here we want to discuss the details of the result reported in
Eq.~(\ref{index_1d}), i.e., that in the case  of the one-dimensional
$XY$ model with nearest-neighbor interactions the index of the
critical points equals the number $n_d$ of ``domain walls'' in the
configuration.

First of all, let us notice that in the present case the Hessian
matrix $H$ is tridiagonal, i.e., it can be written as
\begin{equation}
H = \left(
\begin{array}{cccccc}
\alpha_1 & \beta_1 & 0 & 0 & \ldots & 0 \\
\beta_1 & \alpha_2 & \beta_2 & 0 & \ldots & 0 \\
0 & \ddots & \ddots & \ddots & \ddots & \vdots \\ 
\vdots & \ddots & \ddots & \ddots & \ddots & 0 \\
0 & \ldots & 0 & \beta_{N-2} & \alpha_{N-1} & \beta_{N-1} \\
0 & \ldots & 0 & 0 & \beta_{N-1} & \alpha_N 
\end{array}
\right)
\label{tridiagonal}
\end{equation}
where, assuming free boundary conditions,
\begin{subequations}
\begin{eqnarray}
\alpha_1 & = & \cos (\varphi_{2} - \varphi_{1}) + h \cos
(\varphi_{1})~; \\
\alpha_i & = & \cos (\varphi_{i+1} - \varphi_{i}) + \cos (\varphi_{i} -
\varphi_{i-1}) + h \cos (\varphi_{i})~, \qquad i = 2,\ldots,N-1 ~;\\
\alpha_N & = & \cos (\varphi_{N} -
\varphi_{N-1}) + h \cos (\varphi_{N})~,
\end{eqnarray}
\end{subequations}
and 
\begin{equation}
\beta_i = - \cos (\varphi_{i+1} - \varphi_{i}) ~, \qquad i = 1,\ldots,N-1 ~.
\end{equation}
Since at critical points $\varphi_i = 0$ or $\pi$, we have that for
any $i$ and for any critical point
\begin{equation}
\beta_i = \pm 1
\label{beta}
\end{equation}
while the diagonal elements $\alpha_i$ are 
\begin{subequations}
\begin{eqnarray}
\alpha_1 & = & 1 \pm  h ~; \\
\alpha_i & = & 2 \pm  h ~, \qquad i = 2,\ldots,N-1 ~;\\
\alpha_N & = & 1 \pm h~,
\end{eqnarray}
\label{alpha}
\end{subequations}
if there are no domain walls, i.e., if $n_d = 0$, while they can assume
also the values $\pm h$ and $-2 \pm h$ ($-1 \pm h$ if $i = 1$ or $i =
N$) if $n_d \not= 0$, i.e., if there are domain walls.

Let us now prove that $n_d \not= 0$ is a {\em necessary} condition for
the presence of negative eigenvalues of the Hessian, i.e., for a
nonvanishing index of a critical point. To do that, we recall a
theorem due to Gershgorin (see e.g. Ref.~\cite{Wilkinson}), which, in
the simple case of a real symmetric matrix, can be stated as follows:
\begin{theorem}[Gershgorin]
Let $A$ be a real $n\times n$ 
symmetric matrix whose elements are $a_{ij}$, and let
\[
r_i = \sum_{j\not=i} |a_{ij}|~, \qquad i = 1,\ldots,n~;
\]
the eigenvalues of $A$ lie in the intervals
\[
X_i = \{ x \in \mathbb{R} \, : \, |x - a_{ii}| < r_i \}~;
\]
if $m$ of the $X_i$ form a disjoint set, then precisely $m$
eigenvalues (counted with their multiplicity) lie in it.
\end{theorem}
In our case, due to Eq.~(\ref{beta}), at any critical point we have
\begin{subequations}
\begin{eqnarray}
r_1 ~ = ~ r_N & = & 1 ; \\
r_i & = & 2 ~, \qquad i = 2,\ldots,N-1 ~,
\end{eqnarray}
\end{subequations}
so that, if $n_d = 0$ and $h \to 0$, Eqs.~(\ref{alpha}) and
Gershgorin's theorem imply that all
the eigenvalues lie in the interval $|x - 2| < 2$, so that there are
no negative eigenvalues and the index is zero. On the other hand, if
$n_d \not= 0$ and $h \to 0$, then the intervals $X_i$ are either $|x|
< 2$, or $|x +2| < 2$, hence the eigenvalues lie in the interval $(-4,2)$, so
that the index can be nonvanishing. However, Gershgorin's theorem is
useless to compute the number of negative eigenvalues, because the
intervals $X_i$ overlap each other, thus the eigenvalues cannot be
localized more strictly.

Anyway, the fact that the Hessian is tridiagonal allows us to compute
directly its characteristic polynomial $\det(H - \lambda
I)$, whose roots $\lambda_1,\ldots,\lambda_N$ are the
eigenvalues, by means of a recurrence formula. Let
\begin{subequations}
\begin{eqnarray}
p_0(\lambda) & = & 1~;\\
p_1(\lambda) & = & \alpha_1 - \lambda~; \\
p_k(\lambda) & = &  (\alpha_k - \lambda)p_{k-1}(\lambda)
- \beta^2_{k-1}p_{k-2}(\lambda)
\end{eqnarray}
\label{recurrence}
\end{subequations}
then, since 
\begin{equation}
p_k(\lambda)  =  \det \left(
\begin{array}{cccccc}
\alpha_1 - \lambda & \beta_1 & 0 & 0 & \ldots & 0 \\
\beta_1 & \alpha_2 -\lambda & \beta_2 & 0 & \ldots & 0 \\
0 & \ddots & \ddots & \ddots & \ddots & \vdots \\ 
\vdots & \ddots & \ddots & \ddots & \ddots & 0 \\
0 & \ldots & 0 & \beta_{k-2} & \alpha_{k-1} - \lambda & \beta_{k-1} \\
0 & \ldots & 0 & 0 & \beta_{k-1} & \alpha_k - \lambda
\end{array}
\right)~,\qquad k = 2,\ldots,N~,
\end{equation}
the characteristic polynomial of $H$ is given by
$p_N(\lambda)$. Since at the critical points all
the $\beta$'s are $\pm 1$ -- see Eqs.~(\ref{beta}), we have that
\begin{subequations}
\begin{eqnarray}
p_0(\lambda) & = & 1~;\\
p_1(\lambda) & = & \alpha_1 - \lambda~; \\
p_k(\lambda) & = &  (\alpha_k - \lambda)p_{k-1}(\lambda)
- p_{k-2}(\lambda)
\end{eqnarray}
\end{subequations}
so that the characteristic polynomial $p_N(\lambda)$ depends only on
the $\alpha$'s. Moreover, the following theorem holds (see 
e.g.~Ref.~\cite{matrixbook}):
\begin{theorem}
Let $H$ be a tridiagonal symmetric matrix defined as in
Eq.~(\ref{tridiagonal}). Define the
sequence
\begin{equation}
\{p_0(\lambda),p_1(\lambda),\ldots,p_N(\lambda)\}
\end{equation}
as in Eq.~(\ref{recurrence}); then the number of sign changes in the
sequence (with the rule that if $p_i(\lambda) =0$ then it has the opposite sign
of $p_{i-1}(\lambda)$) equals the number of eigenvalues of  $H$ which
are less than or equal to $\lambda$. 
\end{theorem}
Then the number $n_c$ of sign changes in the sequence 
\begin{equation}
\{p_0(0),p_1(0),\ldots,p_N(0)\}
\end{equation}
equals the number of negative eigenvalues, i.e., the 
index of the critical point because no eigenvalues are zero. If one
puts $h=0$, then there is one eigenvalues which becomes zero at any
critical point, so that the index equals $n_c - 1$, but in this case
one easily sees by direct computation (which can be performed exactly
on a computer at any $N$ 
because in this case the $\alpha$'s are integer numbers)
that $n_d = n_c - 1$, so that one finds the result reported in
Eq.~(\ref{index_1d}).

\newpage

\end{document}